
\magnification=\magstep1
\input paper
\pretolerance=10000
\voffset=-.5truein
\baselineskip=14pt

\def\s{\smallskip}
\def\m{\medskip}
\def\b{\bigskip}
\def\n{\noindent}
\def\c{\centerline}

\line{\hfil IUCAA-19/93 July'93}

\vskip 2 cm

\c{\mid Singularity Free Spacetimes -- I :}
\c{\mid Metric and Fluid Models}

\vskip 1.5 cm
\c{Naresh Dadhich\footnote{$^\dagger$}{E-mail : naresh@iucaa.ernet.in},
L. K. Patel\footnote{$^*$}{Permanent Address~: Department of Mathematics,
Gujarat University, Ahmedabad - 380 009, India} and R.
Tikekar\footnote{$^{**}$}{Permanent
address~: Department of Mathematics, Sardar Patel University,
Vallabh--Vidyanagar
358 120, India}}
\vskip 1cm

\c{Inter University Centre for Astronomy and Astrophysics,}
\c{Post Bag 4, Ganeshkhind, Pune - 411 007.}

\vskip .7 cm

\c{\bf Abstract}
\b\m

We show that the metric for the singularity free family of
fluid models [3] can be obtained by a simple and natural
inhomogenisatiat the oon and anisotropisation procedure from
Friedman--Robertson--Walker metric with negative curvature.
The metric is unique for cylindrically symmetric spacetime
with metric potentials being separable functions of radial
and time coordinates. It turns out that fluid models separate
out into two classes, with
$\rho \not= \mu p$ in general but $\rho = 3p$ in particular
and $\rho = p$. It is shown that in both the cases radial heat
flow can be incorporated without disturbing the singularity
free character of the spacetime. Further by introducing massless
scalar field it is possible to open out a narrow window for
$\mu, 4 \geq \mu \geq 3$.

\vfill
\n PACS numbers~: 04.20Jb, 98.80Dr

\eject

\n{\bf I. Introduction}
\m
The standard view, the universe had a big-bang singular origin
predicted by the standard Friedman--Robertson--Walker (FRW)
model, was supported on the formal grounds by the strong
si
if we adhere to reasonable energy and
causality conditions and to general relativity (GR). To handle the
cosmic singularity several attempts have been made by several
authors. They include modification of GR, invoking quantum
effects and new fields, ascribing unusual properties of matter etc.
By sacrificing energy conditions and other physical
properties, it is easy to construct a singularity free model,
the most well-known example of this class is the deSitter universe.
That is there exists a large number of such models where acceptable
physical behaviour of matter has been traded off for avoidance of
singularity.

Under this background Senovilla's singularity free solution[2]
is remarkable that it satisfies the energy conditions and has very
acceptable equation of state $\rho = 3p$. This is the first solution
of its kind that is free of singularity with physically reasonable
behaviour of matter. Further Ruiz and Senovilla [3] have discussed
the singularity free character of inhomogeneous cylindrically symmetric
spacetime and have separated out a family of singularity free perfect
fluid models. It has been shown [4] that these solutions
are geodetically complete and are able to escape the singularity
theorems because they do not obey the assumption of existence of
causal trapped surfaces.

In this paper I we first establish a linkage between the FRW
metric and the Ruiz--Senovilla singularity free family.
We argue that a natural and simple inhomogenisation and anisotropisation
of the FRW open model leads to the singularity free family.
Transform FRW metric into the cylindrical coordinates, then
inhomogenise and anisotropise it by using the functions of $r$ and $t$ as
they occur in the transformed form and the resulting metric is the
singularity free family. Further the singularity free family is
unique for separation of variables in the cylindrically symmetric
metric potentials. It turns out that fluid solutions separate out
into two classes, having $\rho \not= \mu p$ in general
(but in particular $\rho = 3p$) and $\rho = p$. The stiff fluid solution
yields two distinct cosmological vacuum solutions as its matter
free ($\rho = 0$) limit [5].

 From the point of view of general study of singularity free cosmology
it would be of relevance to examine how robust is the singularity
free character of the spacetime? That is, can we generate singularity free
vacuum solutions or can we introduce heat flow, viscosity etc. retaining the
singularity free character? In this context we have [6] considered viscous
fluid and found that it is not possible to have viscosity non-negative
for all times while heat flow [7] can be incorporated easily. We give
the heat flow generalizations of the two classes of fluid models mentioned
above. Further if we introduce massless scalar field with the fluid
distribution, then the resulting fluid can have an equation of
state $\rho = \mu p$, with $4 \geq \mu \geq 3$.

In $\S$II we describe the inhomogenisation and anisotropisation procedure
of FRW metric that leads to the singularity free family of spacetime.
We also consider the regularity of Weyl and Ricci curvatures for
this metric. $\S$III is devoted to the study of perfect fluid, fluid with
radial heat flow and fluid with massless scalar field models. We conclude
in $\S$IV with discussion of results obtained in the paper.

In the following paper II a general study of the geodesics of the
singularity free metric is carried out and they are shown to be complete
indicating the absence of singularity of any kind. We also explicitly
exhibit the non-occurrence of compact surfaces that makes the singularity
theorems inapplicable to this family of metrics.

\b
\n{\bf II. Singularity Free Metric}
\m
Ruiz and Senovilla [3] have separated out a singularity free
family of solutions of Einstein's equations for perfect fluid.
The spacetime has cylindrical symmetry. With metric potentials
as separable functions of $r$ and $t$ the family is unique.

We begin with the general form of the metric

$$ds^2 = A^2 (dt^2 - dr^2) - B^2  dz^2 - C^2 d\phi^2.\eqno(2.1)$$

\n Here we have $A = A_1 (t) A_2 (r), B = B_1 (t) B_2 (r),$
$C = C_1 (t) C_2 (r)$ and have used the co-ordinate freedom
to write $G_{tt} = \vert g_{rr} \vert$. The metric admits
two space-like Killing vectors ${\partial \over \partial z}$
and ${\partial \over \partial \phi}$ which are mutually as well as
hypersurface orthogonal implying cylindrical symmetry.
That means we should write $C_2 (r) = rf (r)$ so as to satisfy
the regularity conditions on the axis $r = 0$ and the elementary
flatness condition in its vicinity.

The prime requirement for a metric to be singularity free is that
its coefficients have no zeroes for real values of the co-ordinates
and Ricci and Weyl tensors are regular everywhere.
Ricci curvatures for the metric (2.1) are given in the Appendix
and they will be regular provided ${\dot A \over A}, {A' \over A}$
etc. are regular. At the outset $C_2 (r) = rf (r)$ would imply
$A_1 (t) = C_1 (t)$ (by demanding~: coefficient of $r^{-1} = 0$.)
This regularity condition has no reference to matter distribution.

Assuming the matter content of the space-time to be perfect fluid
we impose the fluid conditions, $R_{14} = 0$, and $R_{11} =
R_{22} = R_{33}$ for the comoving velocity field

$${\bf u} = A dt\eqno(2.2)$$

After some calculations it can be shown that $R_{(4)} = 0$
implies that $A_1 (t) = C_1 (t) = T^\alpha (t), B_1 (t) = T^\beta (t)$
and $A_2 (r) = B^q_2 (r)$. That is the time variation in the metric
stems from one single function. Subsequently $R_{11} = R_{22}
= R_{33}$ (from Appendix) will imply

$${\ddot T \over T} + (\alpha + \beta - 1) {\dot T^2 \over T^2}
= k^2 \eqno(2.3)$$

$${C''_2 \over C_2} - {B''_2 \over B^2} = (\alpha - \beta)
k^2 \eqno(2.4)$$

$$C_2 = B'_2 B_2^{(\alpha - \beta) / \beta}.\eqno(2.5)$$

\n Eq. (2.3) admits the general singularity free solution

$$T (t) = K_1 [ C (\lambda kt) + t_0)]^{1/\lambda^2} = C^{1/\lambda^2}
(\lambda kt), \eqno(2.6)$$

\n where we have chosen the integration constants $K_1 = 1, t_0 = 0$
and written $\alpha + \beta = \lambda^2, C (x) = cosh (x)$. The
other solutions for $\lambda^2 \leq 0$ have been ruled out by the
singularity free and positivity of density conditions.

In [3] it was assumed that $\alpha + \beta = 1$ and the time dependence
steming from the same function $T (t)$ which we have shown
above as to follow from $R_{14} = 0$. It turns out that there is
no loss of generality in the former assumption either for it
would just imply an overall scaling in the expressions for density
and pressure (i.e. $C^\alpha (kt) \rightarrow C^{\alpha / \lambda^2}
(\lambda kt)$). We can hence take $\alpha + \beta = \lambda^2 = 1$.
For space-dependance Ruiz and Senovilla employ four arbitrary functions
$(g_{44} \not= \vert g_{11} \vert)$ and obtain the general solution
of the system of equations (2.4) and (2.5) with the condition
$\alpha + \beta = 1$. The solution in our notation, will read as

$$MX' = C_2 / B_2, B_2 = X^{(1-\alpha) / (2\alpha - 1)}$$

$$C_2^2 = B_2^{1/2} [(2\alpha -1)^2 k^2 M^2 X^2 + NX^{(4\alpha -3)/
(2\alpha -1)} -L ]$$

\n where $X = X(r)$ and $M, N, L$ are constants of integration.
In Ruiz-Senovilla case $X (r)$ is arbitrary which gets determined
for us because of the choice $g_{44} = \vert g_{11} \vert$. It turns out
that we can integrate for $X$ in terms of elementary regular functions,
only for $N = 0$ to give the general solution

$$B_2 = C^b (mr), C_2 = m^{-1} S (mr) C^c (mr)$$

\n where $S (mr) = sinh (mr)$ and $A_2 = B_2^q = C^a (mr)$.
The fluid consistency equations (2.4), (2.5) and the equation
$R_{14} = 0$ will give relations between the parameters $a, b, c, m, k$
and $\alpha$.

Thus we obtain the metric for singularity free space-time as

$$\eqalignno{ds^2 = &C^{2\alpha} (kt) C^{2a} (mr) (dt^2 - dr^2) - C^{2\beta}
(kt) C^{2b} (mr) dz^2\cr
&- m^{-2} S^2 (mr) C^{2c} (mr) C^{2\alpha} (kt) d\phi^2&(2.7)\cr}$$

\n which is unique for perfect fluid distribution with appropriate
relations between free parameters.

We next argue that the above metric
can be deduced through a natural anisotropization and inhomogeneisation
of the FRW metric. Consider the FRW metric with negative
curvature,

$$ds^2 = dt^2 - T^2 (t) \bigg( {dr^2 \over 1 + r^2} + r^2 d\theta^2
+ r^2 sin^2 \theta d \phi^2\bigg)\eqno(2.8)$$

\n and transform it into cylindrical co-ordinates

$$ds^2 = dt^2 - T^2 (t) \bigg( {d\bar r^2 \over 1 + \bar r^2}
+ (1 + \bar r^2) dz^2 + \bar r^2 d\phi^2\bigg)\eqno(2.9)$$

\n by the transformation

$$r = (S^2 (z) + \bar r^2 C^2 (z))^{1/2}, tan\theta = {\bar r \over
S(z) \sqrt{1 + \bar r^2}}.\eqno(2.10)$$

\n Further writing $m \bar r = S (m\hat r)$ and then dropping caps
we get

$$ds^2 = dt^2 - T^2 (t) (dr^2 + C^2 (mr) dz^2 + m^{-2} S^2 (mr) d \phi^2).
\eqno(2.11)$$

\n Now inhomogenise and anisotropise by taking different powers of $T (t)$
and $C (mr)$ in the metric coefficients. That is the metric (2.7)
with $T = C (kt)$. Here $m^{-2} S (mr)$ is just ($C_2 (mr) = rf (r)$)
to provide $2\pi$ periodicity to the angular co-ordinate $\phi$ and elementary
flatness near the axis $r = 0$. Hence it does not participate in the
inhomogenisation and anisotropisation
procedure. Thus a natural and simple inhomogenisation and anisotropisation
of the FRW model leads to the
singularity free family given by (2.7).

The metric (2.7) can be written in the spirit of (2.9) as

$$\eqalignno{ds^2 = &(1 + k^2 t^2)^{\alpha - 1} (1 + m^2 r^2)^a
dt^2 - (1 + k^2 t^2)^\alpha (1 + m^2 r^2)^{a-1} dr^2\cr
&- (1 + k^2 t^2)^\beta (1 + m^2 r^2)^b dz^2 - r^2
(1 + k^2 t^2)^\alpha
(1 + m^2 r^2)^c
d \phi^2&(2.12)\cr}$$

\n which goes over to the FRW form for $a = 0 = c, b = 1$ and $\alpha = \beta$.
If one were to look for simple functions that were regular and free
of zeroes, so as to have metric free of singularity, the obvious choice
would have been the hyperbolic function $C (mr)$ or $(1 + k^2 r^2)$.
The amazing thing is that this obvious choice is the right one
giving rise to the unique singularity-free cylindrically symmetric metric
(2.7).

The kinematic parameters for the metric (2.7) expansion $\theta$, shear
$\sigma^2$ and acceleration $\dot u_r$ are given by

$$\theta = (\alpha + 1) k S (kt) C^{-\alpha -1} (kt) C^{-a} (mr)\eqno(2.13)$$

$$\sigma^2 = 2 (2\alpha - 1) k^2 S^2 (kt)C^{-2 (\alpha + 1)}
(kt) C^{-2a} (mr)\eqno(2.14)$$

$$\dot u_r = - am S (mr) C^{-\alpha} (kt) C^{-a -1} (mr).\eqno(2.15)$$

\n They are all regular and finite throughout the spacetime.
The Ricci and Weyl curvatures will also be regular as they involve
$\dot A / A, A' / A$ etc. which are all regular and finite for the
metric (2.7). This would also imply the regular behaviour for the
physical parameters $\rho$ and $p$. There would however be some conditions on
the
parameters.

It is the non-vanishing of acceleration and shear that is responsible
for avoidance of singularity. It is physically conceivable that
acceleration and shear donot let the fluid congruence to focus into a
small-enough a region to form trapped sufraces leading
to singularity. Acceleration of congruence viewed as spatial {\it pressure
gradient} which opposes gravitational attraction and provides the bounce to
transform contraction into expansion at $t = 0$. The shear
makes goedesics of the congruence to slip through without
letting them converge into a small region. It helps in defocussing of the
congruence. It is however obvious that their presence alone
is not sufficient to avoid singularity as we can easily check
by letting $C (kt) \rightarrow S (kt)$ in (2.7). Then the spacetime
is singular at $t = 0$ with acceleration and shear non-zero.
Hence it may be a necessary condition but not sufficient.
For sufficiency we should have regularity of the metric and curvatures.
Then energy conditions will have to be satisfied.

In [8], the geodesics of the metric (2.7)
have been extensively studied and it has been shown that the
above space-time is geodesically complete for $\alpha \geq 0, \alpha
+ \beta \geq 0, \alpha \geq \beta, a \geq 0, a + b \geq 0, a \geq b$
and $b \leq 0$. The geodesic completeness is established without reference
to any matter distribution. It is obvious that the space-time is causally
stable and satisfies stronger causality and energy conditions.
Thus there occurs no singularity in the space-time.

For the metric (2.7) we give below the Ricci and Weyl curvatures to show
explicitly that they are regular. The expressions read as follows:

$$\eqalignno{2 (\psi_0 + \psi_4) A^2 =& m^2 (2a - 3c + b -1)
-k^2 (\alpha - \beta) + 2 \alpha (\alpha - \beta) k^2  T^2 (kt)\cr
&+ m^2 [2a (c-b) + (b-c) (b+c-1)] T^2 (mr)&(2.16)\cr}$$

$$\eqalignno{2 (\psi_0 - \psi_4) A^2 = &mk (\alpha + \beta -1 ) T (kt) T^{-1}
(mr) +\{ 2\alpha (c-b)
\cr
&+ \alpha (2a -b-c) + \beta (b+c-2a) + (b-c)\}
T (mr)&(2.17)\cr}$$

$$\eqalignno{2\psi_2 A^2 = &k^2 (1 - 2\alpha) + m^2 (2a + b - 3c -1)\cr
&+ (b+c-2a+2bc -b^2-c^2)m^2 T^2 (mr)\cr
&+ (2 \alpha + \alpha^2 + \beta^2 - 2\alpha\beta -1) k^2 T^2 (kt)&(2.18)\cr}$$

$$\eqalignno{A^2 R_{11} =&m^2 (1+b+3c) - \alpha k^2 + \{b (b-1) + c (c-1)
- a (b+c+1)\}m^2 T^2 (mr)\cr
&- \{\alpha (\alpha + \beta -1)\} k^2 T^2 (kt)&(2.19)\cr}$$

$$A^2 R_{22} = 2bm^2 - \beta k^2 - \beta(\alpha + \beta -1)
k^2 T^2 (kt) + \{b (b-1) + bc\} m^2 T^2 (mr)\eqno(2.20)$$

$$\eqalignno{A^2 R_{33} =& m^2 (b+3c+1) - \alpha k^2 - \alpha (\alpha
+\beta - 1) k^2 T^2 (kt)\cr
&+ \{c (c-1) + bc\} m^2 T^2 (mr)&(2.21)\cr}$$

$$\eqalignno{A^2 R_{44} =& (2\alpha + \beta) k^2 - 2 am^2 + a (1-b-c) m^2 T^2
(mr)\cr
&+ \{\beta (\beta -1) + \alpha (\alpha - 1) - \alpha (\alpha + \beta + 1)\}
k^2 T^2 (kt)&(2.22)\cr}$$

$$A^2 R_{14} = mk [\beta (b-a) - \alpha (a+b)] T (kt) T (mr)\eqno(2.23)$$

\n where

$$A^2 = C^{2\alpha} (kt) C^{2a} (mr), T (x) = tanh (x).\eqno(2.24)$$

\n The above quantities will be singularity free if the coefficients of
$T^{-1} (mr)$ vanish, so we obtain the condition

$$\alpha + \beta = 1.\eqno(2.25)$$

\n With $\alpha, \beta$ as restrained above, Ricci and Weyl curvatures
are regular and finite everywhere indicating that the metric is free
of any kind of singularity. Now we can introduce matter to determine
other parameters occurring in $R_{ik}$ consistent with the energy conditions.

The conditions for perfect fluid $R_{14} = 0$ and $R_{(11)} = R_{(22)}
= R_{(33)}$ would imply

$$(\alpha - \beta) (\alpha + \beta - 1) = 0\eqno(2.26)$$

$$(\beta - \alpha) k^2 = (b - 1 - 3c) m^2 \eqno(2.27)$$

$$(b-c) (b+c-1) = 0\eqno(.2.28)$$

$$a (b+c+1) = b (b-1-c)\eqno(2.29)$$

$$\alpha (a+b) = \beta (b-a).\eqno(2.30)$$

\n The first condition is satisfied in view of (2.25) which has to
be obeyed always as it is dictated by the regularity of Weyl tensor.

The remaining conditions give rise to the following two cases

\s
\item{(i)} $b = c$
\item{(ii)} $b + c = 1.$

\s
\n These are the only two possibilities for singularity free fluid
models satisfying the energy conditions. The former in general
represents a fluid without an equation of state. Only when
$b = -1/3$, it represents Senovilla's [2] radiation model with
$\rho = 3p$. The latter will always have $\rho = p$, representing
stiff fluid model [5]. In this case when we put $\rho = 0$, we get
two distinct vacuum solutions as its matter free limit [5]. The point
to be noted is that these are all general unique solutions for the
singularity free form (2.7). It is also interesting that the two
cases $\rho \not= \mu p$ in general (but $\rho = {1\over 3} p$ in
particular) and $\rho = p$ separate out nicely.

\b
\n{\bf III Fluid Models}
\m
We shall consider the perfect fluid, fluid with radial heat flux
and fluid with massless scalar field models. It is possible to introduce
heat flux in both the cases considered above. Inclusion of massless
scalar field opens out a narrow window $4 \geq \mu \geq 3$ for
$\rho = \mu p$ which otherwise had only the discrete value $\mu = 3$
allowed in the case (i).

\m
\n{\it 3.1 Perfect fluid}
\s
Einstein's equations for non-empty spacetime are

$$R_{ik} = - 8 \pi ( T_{ik} - {1\over 2} T g_{ik})\eqno(3.1)$$

\n where for perfect fluid

$$T_{ik} = (p + \rho) u_i u_k - p g_{ik}, u_i u^i = 1 .\eqno(3.2)$$

\m
\n{\it Case (i) $b = c$}.

\n We write from eqs. (2.27)--(2.30)
\s
$$\alpha = (1 + b) (1 + 2b)^{-1}, a = -b (1+2b)^{-1},
{m \over k} = n = (1 + 2b)^{-1}.\eqno(3.3)$$

\n This determines the metric (2.7) in terms of the two free
parameters $b$ and $k$, say.

The density and pressure have the expressions~:

$$8\pi p A^2 = {b^2 (1-2b) \over (1 + 2b)^3} k^2 C^{-2} (nkr)
- {(3b + 1) (b + 1) \over (1+2b)^2} k^2 C^{-2} (kt)\eqno(3.4)$$

$$8\pi p A^2 = {b(2b-1) (2+3b)\over (1+2b)^3} k^2 C^{-2} (nkr) -
{(3b + 1) (b+1) \over (1 + 2b)^2}
k^2 C^{-2} (kt)\eqno(3.5)$$

\n where $A = C^\alpha (kt) C^a (nkr)$. This is the Ruiz-Senovilla
[3] model. For satisfaction of the energy conditions wrequire $- {1\over 2} < b
\leq - {1\over 3}$ and the Senovilla
radiation model results for $b = - {1\over 3}$. With $b$
so restricted, $\rho$ and $p$ are positive for entire spacetime.
For $b \not= - {1\over 3}$, there is no equation of state of
the type $\rho = \mu p$. The maximum density would occur at
$t = 0$ and $r =0$ which decreases to zero as $t \rightarrow \pm \infty$
and $r \rightarrow \infty$.
The parameter $k$ can be identified with $\rho_{max}$ which can be
freely chosen. At a given $t$, $\rho$ is largest at the origin
$r = 0$ and at given $r$, it is largest at $t = 0$.

\m
\n{\it Case (ii) $b + c = 1$.}
\s
\n In this case we have

$$\alpha = {1\over 2} (2 - b), a = b (b - 1), c = 1 - b, n^2 =
{m^2 \over k^2} = {1\over 4}\eqno(3.6)$$

\n and we get the stiff fluid [5] model with

$$8 \pi \rho = 8 \pi p = \bigg({b^2 - 4 \over 4 A^2} \bigg)
k^2 C^{-2} (kt). \eqno(3.7)$$

\n Clearly the energy condition will be obeyed for $b^2 \geq 4$, i.e.
$b$ lying outside the interval $-2 \leq b \leq 2$, the end points of which
give the two distinct vacuum cosmological solutions [5]. These
solutions are given as follows:

\n For $b = 2$

$$ds^2 = C^4 (mr) (dt^2 - dr^2 - C^2 (2mt) dz^2) - m^{-2} S^2
(mr) C^{-2} (mr) d\phi^2.\eqno(3.8)$$

\n For $b = -2$

$$\eqalignno{ds^2 = &C^4 (2mt) [C^{12} (mr) (dt^2 - dr^2)
- C^6 (mr) m^{-2} S^2 (mr) d\phi^2]\cr
&- C^{-2} (2mt) C^{-4} (mr) dz^2.&(3.9)\cr}$$

\n It can be easily verified from eqs. (2.16)--(2.18) that both
the solutions are Petrov type I. The former (3.6) is static as Weyl scalars
have no time dependence, while the latter can asymptotially represent
a plane gravitational wave. Since there are no obvious localized
sources for the solutions, hence the only possible source can
be gravitational radiation. However, one knows that in realistic situation
the metric for gravitational
[7].

\m
\n{\it 3.2 Fluid with heat flow}
\s
The energy momentum tensor for fluid with heat flow is given by

$$T_{ik} = (p + \rho) u_i u_k - p g_{ik} + (q_i u_k + q_k u_i),
u_i u^i = 1, u_i q^i = 0.\eqno(3.10)$$

\n Here $q_i$ is the heat flow vector. Here we are relaxing the condition
(2.30) by taking radial heat flow. We take the tetrad components of
$q_i$ as $q_{(a)} = (q, 0, 0, 0)$. The parameters $\rho, p$ and $q$ are
given by

$$\eqalignno{8\pi \rho A^2 = &(a-3b) m^2 + (\beta - \alpha) k^2
+ (\beta^2 + 2\alpha \beta + \alpha - \beta) k^2 T^2 (kt)\cr
&+ {1\over 2} (a - 3b) (b+c-1) m^2 T^2 (mr)&(3.11)\cr}$$

$$\eqalignno{8\pi p A^2 = &(a+b) m^2 - (\alpha + \beta) k^2 +
(\alpha + \beta - \beta^2) k^2 T^2 (kt) \cr
&+ {1\over 2} (a+b) (b+c-1) m^2 T^2 (mr)&(3.12)\cr}$$

$$8\pi q A^2 = mk [\alpha (a+b) -\beta (b-a)]
T(kt) T (mr)\eqno(3.13)$$

\n where $A = C^\alpha (kt) C^a (mr).$

\m

\n{\it Case (i) $b = c$.}
\s
\n In this case we have

$$b = c, \alpha = 1 - \beta, a = - {b \over 1 + 2b},
n^2 = {m^2 \over k^2} = {1-2\beta \over 1 + 2b}\eqno(3.14)$$

\n We treat $b$ and $n$ as arbitrary parameters. The physical
parameters are then given by

$$8 \pi q A^2 = {k^2 nb \over (1+2b)} \{ (n^2 - 1) + 4b
(b+1) n^2 \} T(kt) T (nkr)\eqno(3.15)$$

$$\eqalignno{8\pi p A^2 = & {k^2 \over 4} (n^2 -1) \{
1 - n^2 (1+2b)^2\} + {b^2 (1-2b) \over (1+2b)}
n^2 k^2 C^{-2} (nkr)\cr
&+ {1\over 4} \{n^2 (1+2b) -3\} \{1 + n^2 (1+2b)\} C^{-2}
(kt)&(3.16)\cr}$$

$$\eqalignno{8\pi \rho A^2 = &{k^2 \over 4} (3 + n^2)
\{1 - n^2 (1+2b)^2\}\cr
&- {b (1-2b) (2+3b) \over (1+2b)} n^2 k^2 C^{-2} (nkr)\cr
&+ {1\over 4} \{ n^2 (1+2b) -3\} \{1 + n^2 (1+2b)\} C^{-2}
(kt).&(3.17)\cr}$$

\n From the above equations it can be easily seen that the physical
requirements $p \geq 0, \rho \geq 0$ are satisfied for

$$-{1\over 2} < b \leq - {1\over 3},~~~~~~~~~~{3\over 1+2b} < n^2
\leq {1 \over (1+2b)^2}.\eqno(3.18)$$

\n From (3.15) it is clear that $q = 0$ implies $n^2 = (1 + 2b)^{-2}$
and the model reduces to the perfect fluid (3.3). For $n = 3$ we have the
heat flow generalization of the Senovilla radiation model [7].

\m
\n{\it Case (ii) $b + c = 1$}
\s
\n Here we write

$$a  = b (b - 1), c = 1 - b, \alpha = 1 - \beta, n^2 = {m^2 \over k^2}
= {2 \beta -1 \over 4 (b -1)} \eqno(3.19)$$

\n and obtain

$$8\pi q A^2 = b (b -1) (4 n^2 - 1) nk^2 T (kt) T (nkr)\eqno(3.20)$$

$$\eqalignno{8\pi p A^2 = &{k^2 \over 4} (4 n^2 - 1) \{ 1 - 4n^2 (b-1)^2\}\cr
&+ {k^2 \over 4} [ 4n^2 (b-1) -1] [4n^2 (b-1) + 3] C^{-2} (kt)
&(3.21)\cr}$$

$$\eqalignno{8\pi \rho A^2 = &-{k^2 \over 4} [ 4n^2 -1] [3 + 4n^2 (b-1)^2]\cr
&+{k^2 \over 4} [4n^2 (b-1) -1] [4n^2 (b-1) + 3] C^{-2} (kt).&(3.22)\cr}$$

\n The physical requirements $\rho > 0$ and $p \geq 0$ restrict
$n$ and $b$ to the ranges

$$0 < n^2 \leq {1\over 4}, ~~~~~~~~~~1 < b < 2.\eqno(3.23)$$

\n For $n^2 = {1\over 4}$, it reduces to the stiff fluid model (3.6).

The phenomenological expression for the heat conduction is given by

$$q_k = \psi \bigg({\partial F \over \partial x^i} + F \dot u_i\bigg)
(\delta^i_k
- k^i u_k)\eqno(3.24)$$

\n where $\psi$ is the thermal conductivity and $F$ is the temperature.
For the cases under consideration only radial heat flow is retained.
Hence eq. (3.24) can be integrated if $\psi$ is a function of $t$
alone to give

$$\eqalignno{F = & l (t) C^a (nkr)\cr
&- {k \over 16 \pi a \psi} \{
\alpha (a+b) - \beta (b-a)\} C^{-a} (nkr) C^{-\alpha} (kt) T (kt)&(3.25)\cr}$$

\n where $l (t)$ is an arbitrary function of $t$. Thus temperature
distributions for the cases (i) and (ii) can be obtained.

The heat flux has been considered in the evolution of the cosmologial
models by several authors (see [7]).

\m
\n{\it 3.3. Fluid with massless scalar field}
\s
\n The energy momentum tensor for a perfect fluid with massless
scalar fields $V$ is given by

$$T_{ik} = (\rho + p) u_i u_k - p g_{ik} + V_i V_k - {1\over 2}
g_{ik} V^a V_a \eqno(3.26)$$

\n where $u^i u_i = 1$ and $V_i = {\partial V \over \partial x^i}$.
We take the scalar field $V$ to be function of $t$ only.
$V$ has to satisfy the equation

$$[\sqrt{-g} V^i],_i = 0\eqno(3.27)$$

\n the first integral of which is given by

$$\dot V = lC^{-(\beta + \gamma)} (kt)\eqno(3.28)$$

\n where $l$ is a constant of integration. The introduction
of the scalar field does not affect the fluid conditions and hence
the free parameters continue to be governed by the conditions (2.26)--(2.30).
In this case also the solution separate into two classes,
(i) $b = c$ and (ii) $b + c = 1$. In the latter case the solution
represents a stiff fluid in the presence of scalar field. Since any stiff fluid
solution also has a scalar field interpretation this case does not lead
to any new possibility. However, in the former case $b = c$, we obtain

$$\alpha = {1 + b \over 1 + 2b}, ~~~~~a = {-b \over 1 + 2b},~~~~~
n^2 = {m^2 \over k^2} = (1 + 2b)^{-2}, ~~~~~\beta = 1 - \alpha.\eqno(3.29)$$

\n The density and pressure have the expressions~:

$$\eqalignno{8 \pi \rho A^2 = & {b (2b - 1) (2 + 3b) \over (1 + 2b)^3}k^2
C^{-2}
(nkr)\cr
&- \bigg[{(b +1) (3b +1) \over (1 + 2b)^2} k^2 + l^2 \bigg]
C^{-2} (kt)&(3.30)\cr}$$

$$\eqalignno{8\pi p A^2 = &{b^2 (1 - 2b) \over (1 + 2b)^3} k^2 C^{-2} (nkr)\cr
&- \bigg[ {(b+1) (3b+1)\over (1+2b)^2} k^2 + l^2 \bigg] C^{-2}
(kt)&(3.31)\cr}$$

\n where $A = C^\alpha (kt) C^a (nkr).$

\n For $b = -{1\over 3}$, the solution represents a scalar-field
generalization of Senovilla radiation model.
For $l^2 = - {(b+1) (3b+1) \over (1+2b)^2} k^2$ the fluid distribution
of the model has an equation of state $\rho = \mu p$ where $\mu = -{1\over b}
(2+3b)$. The positivity of $\rho$ and $p$ restricts $b$ to the range
$-{1\over 2} < b \leq -{1\over 3}$ so that $3 \leq \mu \leq 4$.

\b

\n{\bf IV. Discussion}
\m
We have argued that the singularity free family of perfect
fluid spacetimes obtained by Ruiz and Senovilla [3]
can be inferred through a simple and natural inhomogenisation
and anisotropisation of FRW metric for the open universe. This
establishes a kind of linkage between the two. If one were to
write a singularity free metric, without reference to anything
else, the natural choice for metric functions would have been
hyperbolic or quadratic functions without zeros. The amazing
thing is that this obvious choice is the right and the only
choice. However, one has to assume cylindrical symmetry in place
of spherical symmetry.

It turns out that the metric (2.7) is the unique
singularity free spacetime for cylindrically symmetric spacetime
with separable variables. In the context of cosmology separability
is an appropriate assumption for the overall behaviour of universe
should not depend upon the interactions arising out of interaction
between variables. We are already taking a big step deeper by giving
up homogeneity and isotropy of the today's universe as described
by FRW metric. This was motivated by the consideration that universe
has to have inhomogeneity and anisotropy at early times so as to
have generic initial conditions as well as to facilitate
formation of structures. It is obvious that in order to consider
anisotropy and inhomogeneity effectively one has to consider less
restrictive symmetry than spherical. That leads naturally to
cylindrical symmetry. We would hence argue that choice of cylindrical
symmetry is almost determined by inhomogeneity and anisotropy of the
early universe. There can, however, be more general metric without
symmetry but cylindrical symmetry is the simplest choice.

For the avoidance of singularity acceleration and shear play the crucial
role. The former provides the bounce to the universe at $t = 0$
where contraction turns into expansion while the latter makes the
fluid geodesic congruence to slip through without letting them to
focus in a small enough region to form compact surfaces leading
to singularity. The overall scenario is~: universe has very low
density tending to zero at $t \rightarrow - \infty$, wherefrom it starts
contracting and attains the dense state at $t = 0$ where contraction
changes to expansion and it expands to low density again at
$t \rightarrow \infty$. The momentum gained during the contraction phase
takes the universe through $t = 0$ to the expanding phase. The maximum
density at $t = 0$ and $r = 0$ can be made as large or small as one
pleases by choosing the parameter $k$. All through all the physical and
kinematic parameters remain regular and finite.
The presence of acceleration and shear may be necessary
but by no means sufficient to avoid singularity. In addition the spacetime
should be regular.

Let us recall that $\alpha = \gamma$ was dictated by the cylindrical symmetry
while $\alpha + \beta = 1$ for the metric (2.7) is demanded by the Weyl
regularity. These two are general conditions without reference to
any matter field. That is of the three $\alpha, \beta, \gamma$ only one is
free for the singularity free spacetime. It may be noted that all the fluid
models (which are the only ones for the metric (2.7)) discussed in
$\S$III obey the conditions, $\alpha \geq 0, \alpha + \beta \geq 0, \alpha \geq
\beta, a \geq 0, a + b \geq 0,
a \geq b$ and $b \leq 0$, obtained in the following paper II
for completteness of geodesics.

The most important question for the singularity free models is how to
evolve them into the standard FRW model which successfully describes
the present day universe. The affirmative answer to this question will
bring these models into the active arena of cosmology and would perhaps
have very significant role to play in the early universe cosmology. The main
difficulty here is that the anisotropy measure $(\sigma / \theta)$ of the
metric (2.7) is constant which means it remains anisotropic for all times.
It would be interesting to find a singularity free solution with $\sigma /
\theta$
decreasing with $t$ so that it can isotropise at late times to go over to FRW
model. This brings us to another important question~: do there exist other
singularity free solutions with lesser or no symmetry, or is cylindrical
symmetry singled out? These are the questions that are currently engaging
us.

A brief report of important results discussed here has been submitted for
publication elsewhere [10].

L.K. Patel and R. Tikekar thank IUCAA for hospitality.

\vfill\eject

\n{\bf References}
\b
\item{[1]}S.W. Hawking and G.F.R. ellis, The large scale structure
of space-time (Cambridge University Press, Cambridge, England, 1973).
\s
\item{[2]}J.M.M. Senovilla, Phys. Rev. Lett. {\bf 64}, 2219 (1990).
\s
\item{[3]}E. Ruiz and J.M.M. Senovilla, Phys. Rev. {\bf D45}, 1995 (1992).
\s
\item{[4]}F.J. Chinea, L. Fernandez--Jambrina and J.M.M. Senovilla,
Phys. Rev. {\bf D45}, 481 (1992).
\s
\item{[5]}L.K. Patel and N. Dadhich, Singularity free inhomogeneous
cosmological
stiff fluid models, IUCAA--1/93 -- Preprint (1993).
\s
\item{[6]}L.K. Patel and N. Dadhich, Singularity free inhomogeneous viscous
fluid cosmological models, IUCAA-21/92 -- Preprint (1992).
\s
\item{[7]}L.K. Patel and N. Dadhich, To appear in Class. Quantum Grav.
(1993).
\s
\item{[8]}N. Dadhich, L.K. Patel and R. Tikekar, Paper-II (1993).
\s
\item{[9]}F.A.E. Pirani, In Gravitation~: an introduction to current
research, Ed. L. Witten (John Wieley, New York, 1962).
\s
\item{[10]}N. Dadhich, R. Tikekar and L.K. Patel, On singularity free
cosmological models, IUCAA--16/93 -- Preprint (1993).
\vfill\eject

\n{\bf Appendix}
\b
\n For the metric (2.1) we introduce the tetrad $\theta^1 = A dr,
\theta^2 = B dz, \theta^3 = C d\phi, \theta^4 = A dt$. The
non-vanishing tetrad componentes $R_{ab}$ of the Ricci
tensor for (2.1) are given by

$$R_{14} = {1 \over A^2}
\bigg[{\dot B' \over B} + {\dot C' \over C} - {\dot A \over A}
\bigg({B' \over B} + {C' \over C}\bigg) - {A' \over A}
\bigg({\dot B \over B} + {\dot C \over C}\bigg) \bigg]$$

$$\eqalign{R_{44} = &{1 \over A^2} \bigg[{\ddot A \over A} + {\ddot B \over B}
+ {\ddot C \over C} - {\dot A \over A} \bigg({\dot A \over A}
+ {\dot B \over B} + {\dot C \over C}\bigg)\cr
&- {A'' \over A} + {A' \over A} \bigg({A' \over A} - {B' \over B}
- {C' \over C}\bigg) \bigg],\cr}$$

$$\eqalign{R_{11} = &{1 \over A^2} \bigg[{A'' \over A} + {B'' \over B}
+ {C'' \over C} - {A' \over A} \bigg({A' \over A}
+ {B' \over B} + {C' \over C}\bigg)\cr
&- {\ddot A \over A} - {\dot A \over A} \bigg({\dot B \over B} + {\dot C \over
C}
- {\dot A \over A}\bigg) \bigg],\cr}$$

$$R_{22} = {1 \over A^2} \bigg[{B'' \over B} + {B'C' \over BC} - {\ddot B
\over B} - {\dot B \dot C \over BC}\bigg]$$

$$R_{33}= {1 \over A^2} \bigg[{C'' \over C} + {B'C' \over BC}
- {\ddot C \over C} - {\dot B \dot C \over BC}\bigg]$$

\n A prime and a dot indicate differentiation with respect to $r$
and $t$ respectively.
\bye